\documentclass[11pt]{article}

\newcommand{\be}{\begin{equation}}
\newcommand{\ee}{\end{equation}}
\newcommand{\beq}{\begin{eqnarray}}
\newcommand{\eeq}{\end{eqnarray}}

\newcommand{\tE}{\lefteqn{\smash{\mathop{\vphantom{<}}\limits^{\;\sim}}}E}
\newcommand{\tP}{\lefteqn{\smash{\mathop{\vphantom{<}}\limits^{\;\sim}}}P}
\newcommand{\tQ}{\lefteqn{\smash{\mathop{\vphantom{<}}\limits^{\;\sim}}}Q}
\newcommand{\Et}{\lefteqn{\smash{\mathop{\vphantom{\Bigl(}}\limits_{\sim}
\atop \ }}E}
\newcommand{\Pt}{\lefteqn{\smash{\mathop{\vphantom{\Bigl(}}\limits_{\sim}
\atop \ }}P}
\newcommand{\Qt}{\lefteqn{\smash{\mathop{\vphantom{\Bigl(}}\limits_{\sim}
\atop \ }}Q}

\newcommand{\tN}{\lefteqn{\mathop{\vphantom{\Bigl(}}\limits_{\sim}
\atop \ }N}

\newcommand{\tNn}{\lefteqn{\mathop{\vphantom{'}}\limits_{\sim}}{\cal N}}

\newcommand{\R}{R}
\newcommand{\CS}{{\cal S}}
\newcommand{\SA}{{\cal A}}
\newcommand{\tSA}{{\widetilde \SA}}
\newcommand{\SSA}{{\bf A}}

\newcommand{\iAas}[2]{{A^{\smash{({\rm ash})}#2}_{#1}}}
\newcommand{\Aash}{{A^{\smash{({\rm ash})}}}}
\newcommand{\iAash}[2]{{A^{\smash{({\rm ash})}#2}_{#1}}}
\newcommand{\Abar}{{A^{\smash{({\rm bar})}}}}
\newcommand{\iAbar}[2]{{A^{\smash{({\rm bar})}#2}_{#1}}}
\newcommand{\SSAr}{{\SSA^{\smash{({\rm ash})}}}}
\newcommand{\iSSAr}[2]{{\SSA^{\smash{({\rm ash})}#2}_{#1}}}
\newcommand{\tPb}{{\tP_{\smash{(\im)}}}}

\def \Gc{\G^{(\chi)}}

\newcommand{\nd}{{\cal N}_D}

\newcommand{\im}{\beta}
\newcommand{\Rb}{{\rm \bf R}}
\newcommand{\Cb}{{\rm \bf C}}
\newcommand{\Nat}{{\rm \bf N}}

\newcommand{\G}{{\cal G}}

\newcommand{\CH}{{\cal H}}

\newcommand{\CU}{{\cal U}}

\newcommand{\Ppr}[2]{I_{(#2)}^{(#1)}}

\newcommand{\eps}{\varepsilon}

\newcommand{\p}{\partial}

\newcommand{\ff}{\Lambda}

\newcommand{\IQ}{{I_{\smash{(Q)}}}}

\def\SL{$SL(2,C)$ }
\def\SU{$SU(2)$ }
\newcommand{\sg}{$SU(2)$ }
\newcommand{\sgchi}{$SU_{\chi}(2)$ }
\newcommand{\sgi}[1]{$SU_{#1}(2)$ }

\newcommand{\Ref}[1]{(\ref{#1})}
\def \f{\frac}
\def \tl{\tilde}

\def \sl{SL(2,C)}

\begin{document}
%
%
\title{SU(2) Loop Quantum Gravity seen from Covariant Theory}
\author{
{\bf Sergei Alexandrov}\thanks{e.mail: alexand@spht.saclay.cea.fr.
Also at V.A.~Fock Department of Theoretical Physics, St.~Petersburg
University, Russia} \\
{\small Service de Physique Th{\'e}orique, CNRS - URA 2306,
C.E.A. - Saclay,}\\
{\small F-91191 Gif-sur-Yvette Cedex, France}\\
{\small and }\\
{\small Laboratoire de Physique Th{\'e}orique de l'{\'E}cole Normale
Sup{\'e}rieure,}\\
{\small 24 rue Lhomond, 75231 Paris Cedex 05, France}\\
\\
{\bf Etera R. Livine}\thanks{e.mail: livine@cpt.univ-mrs.fr} \\
{\small Centre de Physique Th{\'e}orique, Campus de Luminy, Case 907,}\\
{\small 13288 Marseille Cedex 09, France}
}
\date{(sept 26, 2002)}
\maketitle

\begin{abstract}
Covariant loop gravity comes out of the canonical analysis of the Palatini
action and the use of the Dirac brackets arising from dealing with the
second class constraints (``simplicity'' constraints). Within this
framework, we underline a quantization ambiguity due to the existence of a
family of possible Lorentz connections. We show the existence of a Lorentz
connection generalizing the Ashtekar-Barbero connection and we
loop-quantize the theory showing that it leads to the usual $SU(2)$ Loop
Quantum Gravity and to the area spectrum given by the $SU(2)$
Casimir. This covariant point of view allows to analyze closely the
drawbacks of the $SU(2)$ formalism: the quantization based on the
(generalized) Ashtekar-Barbero
connection breaks time diffeomorphisms and physical outputs depend
non-trivially on the embedding of the canonical hypersurface into the
space-time manifold. On the other hand, there exists a true space-time
connection, transforming properly under all diffeomorphisms.
We argue that it is this connection that should be used in the definition
of loop variables. However, we are still not able to complete the
quantization program for this connection giving a full solution of the
second class constraints at the Hilbert space level. Nevertheless, we
show how a
canonical quantization of the Dirac brackets at a finite number of points
leads to the kinematical setting of the Barrett-Crane model, with simple
spin networks and
an area spectrum given by the $SL(2,C)$ Casimir.
\end{abstract}

\newpage

\section{Introduction}

Loop Quantum Gravity as developed today seems to be a promising
approach for quantizing general relativity (for reviews see
\cite{rovelli,thiemann}).
Although it gives some interesting results like discrete quanta of area
and volume \cite{area,ALarea}
and a derivation of the black hole entropy \cite{bh},
there appear several problems. First of all, it is based
on the use of a space triad and an \SU connection where \SU
is the gauge group for the three dimensional space.
This particular choice of variables loses
the explicit covariance of the theory and
a space-time geometrical interpretation \cite{Sam}.
Moreover, there exists an additional puzzle:
a free parameter in the theory, the so-called
Immirzi parameter \cite{imm}. This parameter comes
out of a canonical transformation but creates
a full one-parameter family of quantizations
which are not unitarily equivalent \cite{rt}.
It was an open problem to understand the physical
relevance of the Immirzi parameter and how it effectively
influences the dynamics of the quantum theory.
It turned out that this problem can be studied from
a new point of view in the framework of an explicitly covariant
formalism \cite{sergei1}. The obtained results suggest
that the Immirzi parameter should disappear
from the physical output of a path integral formulation of quantum gravity
\cite{sergei1} as well as of its canonical quantization based on
this covariant formulation \cite{sergei2,sergei3}.
The goal of the present paper is to explain
how one can derive the \SU Loop Quantum Gravity (LQG)
from the covariant canonical quantization. This will allow us
to tackle the issues of LQG from this different point of view, and
discuss the drawbacks of LQG.

Loop quantum gravity with the Immirzi parameter was
shown to come from a canonical analysis of
the generalized Hilbert-Palatini action in
the so-called time gauge  \cite{holst,barros}.
An explicitly covariant canonical analysis
of this action was carried out in \cite{sergei1}
and led to a proposal for its quantization
in \cite{sergei3,sergei4}.
Although in \cite{sergei4} a Hilbert space for the quantum theory
has been proposed, it is not clear whether it is the right solution or not.
A rigorous construction of such a space of quantum states remains
to be done and there are still many questions to be answered within this
new formalism.
Besides the issues related to the non-compactness of the Lorentz gauge
group \cite{spinnet},
the situation is complicated by the nontrivial
canonical structure of the theory.
Indeed, since the covariant analysis was done through introducing the Dirac bracket
taking into account the second class constraints (also called simplicity
constraints), the commutation relations of the basic variables
have changed. In particular, the connection becomes non-commutative
which is a major obstacle to
understanding the geometrical meaning of the theory
and to building an appropriate Hilbert space.


Nevertheless, a strong result of the formalism is that there exists
a unique Lorentz connection in the theory which transforms properly
under space-time diffeomorphisms. It is the true space-time
connection. Still, its geometrical interpretation in quantum theory is not
straightforward since it is non-commutative. However,
it turns out possible to write observables and a Hilbert space of
quantum geometry states using functionals of the connection and
of the foliation. This brings the theory close in its formulation
to the spin foam setting \cite{psn}.
Moreover, a nice feature of the resulting theory is that
the physical output of the theory does not depend on the Immirzi
parameter. This quantization seems the most natural since it respects
all the classical symmetry and does not break the space-time
diffeomorphism invariance.

Now, is there a place for the usual \SU Loop Quantum Gravity
in the framework of the covariant theory? The answer is affirmative.
It turns out there exists a natural covariant generalization
of the Ashtekar-Barbero connection what makes possible to derive
LQG starting from the covariant quantization.
Moreover, this connection is the only commutative one.
This last feature simplifies a lot the quantization process.
It turns out to yield exactly the same Hilbert space as \SU LQG
and to reproduce the area spectrum of the \SU approach.
This derivation establishes an exact correspondence between
the covariant formalism and the usual one.
The study of this case is interesting because it can be a guide
for the "correct" diffeomorphism preserving quantization
since it is possible to solve explicitly the second class constraints
at the quantum level using
this generalized Ashtekar-Barbero connection.
It also helps to look at the issues of \SU LQG from
a new point of view since the problems encountered in this new
(covariant) approach are unavoidable in \SU LQG.
In particular, the scalar (Hamiltonian) constraint
turns out hard to understand and the \SU theory
definitively breaks the space-time diffeomorphism invariance,
as it was foreseen in \cite{Sam}.


The paper is organized as follows.
We begin in section 2 by considering the basic features of possible ways
to construct the canonical formulation of general relativity with the
Lorentz gauge group.
We introduce several objects, generalizing the Ashtekar-Barbero
connection in different ways, which are shown to be all related with
each other.
Then we introduce the covariant Ashtekar-Barbero
connection and give a precise account of its properties.
Using this connection, we quantize the theory
following the usual techniques of the loop approach.
Namely, we construct the corresponding Hilbert space and show
that, in a particular gauge, it reproduces the Hilbert space of
the $SU(2)$ approach.
In other words, we derive $SU(2)$ LQG from
Covariant Loop Gravity at the level of the
Hilbert space.
Then we explain different drawbacks of the $SU(2)$ formalism,
in particular we point out that it breaks the diffeomorphism
invariance at the quantum
level. We argue that a correct quantization should be based
on the covariant space-time connection described in section 3.
We also discuss the link of the canonical formalism with the spin foam
approach. Spin foams should arise as models of the space-time
resulting from LQG \cite{RR}.
The current model, the Barrett-Crane model, is shown to have some
similarities with the present covariant approach and shares the similar
kinematical Hilbert space of quantum states.
In section 4, we comment on the role of the Immirzi parameter in Loop
Quantum Gravity.
Section 5 is devoted to conclusions and discussions.

\section{Deriving the $SU(2)$ formalism from the covariant one}

\subsection{Canonical formulations and the Ashtekar--Barbero connection}

\label{ash-bar}

The action for general relativity that we study here is the
generalized Hilbert-Palatini action:
\be
S_{(\beta)}=\frac12 \int \eps_{\alpha\beta\gamma\delta}
e^\alpha \wedge e^\beta \wedge (\Omega^{\gamma\delta}+
\frac{1}{\beta}\star\Omega^{\gamma\delta}), \label{palat}
\ee
where $e^\alpha$ ($\alpha$ is an internal Lorentz
index) is the tetrad field,
$\Omega^{\alpha\beta}$ is the curvature of
the spin-connection $\omega^{\alpha\beta}$ and
the star operator is the Hodge operator defined as
$ \star \Omega^{\alpha\beta}=\frac 12 {\eps^{\alpha\beta}}_{\gamma\delta}
\Omega^{\gamma\delta} $.
Under the restriction that the tetrad $e$ is not degenerate,
the equations of motion of this action lead to the usual
Einstein equations and thus do not modify general relativity. Nevertheless,
the addition of an extra term compared to the
original Palatini action leads to the introduction of a new coupling
constant $\im$.
As was shown by Holst \cite{holst}, in the so-called {\it time gauge},
this additional term leads to Loop Quantum Gravity
with $\im$ as Immirzi parameter in the quantum theory.
Therefore, it was suggested that $\im$ gives rise to a new
fundamental physical constant \cite{rt}.

A summary of the canonical analysis of the action (\ref{palat})
without any gauge fixing can be found in appendix \ref{A}.
There are second class constraints in the theory.
There are two main ways to deal with such a system. We can
either solve them or take them into account in the symplectic
structure by modifying the Poisson bracket to the Dirac bracket.

The first alternative has been worked out by Barros e S{\'a} in
\cite{barros}. After solving the second class constraints, the natural
configuration variables parametrising the system are the
field $\chi^a=-{e^{ta}}/{e^{t0}}$ ($a$ being an $su(2)$ index
and $t$ the $0$ space-time index),
which is the space components of the time normal
or internal time direction, and a generalization of
the Ashtekar-Barbero connection of LQG ($i$ being a space index):
\be
\iAbar{ia}{}=\omega^{(\im)}_{i0a}+\omega^{(\im)}_{iab}\chi^b,
\label{connecbarros}
\ee
where we use the notation
$\omega^{(\im)}_{\alpha\beta}=\omega_{\alpha\beta}
+\f{1}{\im}*\omega_{\alpha\beta}$.
To show the relation of $\Abar$ with
the Ashtekar-Barbero connection, one imposes the time gauge.
In these variables, it is described by the choice
$\chi=0$, which can be achieved by using the Gauss law constraints
generating the internal Lorentz boost transformations.
In this gauge, one finds the exact set-up of LQG
with Immirzi parameter $\im$, reproducing Holst's result \cite{holst}.
In particular, $\im \Abar$ coincides with the usual
Ashtekar-Barbero connection, when expressing $\omega_i^{ab}$
through the triad by means of (a half of) the Gauss constraints.

The disadvantage of this formalism is that it breaks the covariance
of the theory when solving explicitly the second class constraints.
As a result, it becomes rather complicated and awkward for
making calculations.
To simplify the calculations, one imposes the time gauge,
which breaks the boost part of the Lorentz symmetry and returns us to
the usual \SU formulation. Thus, this way can add nothing new
to our understanding.
Besides, the canonical variable $\Abar$ is in fact neither a
Lorentz nor an \SU connection. Therefore, it turns out to be
inappropriate for loop quantization, especially, if one seeks
a quantization preserving the Lorentz symmetry.


The other alternative, followed by one of the author,
is to use the second class constraints to induce
a Dirac bracket \cite{sergei1}. This allows to leave the Lorentz covariance
untouched and thus to construct an explicitly covariant theory.
Doing so, we can keep an \SL connection as a canonical variable.
However, due to the presence of the Dirac bracket,
the canonical variables do not play a preferable role anymore and
there exists many other Lorentz connections
which can be constructed from the canonical variables.
This gives rise to a quantization ambiguity in the loop approach,
since each of them can be used in the definition of loop variables.
Then, following the philosophy of LQG,
if we require that the area operator be diagonal on the Wilson loops
defined by the connection, we end up with a two-parameter family
of possible \SL connections.
Following the methods of LQG, one can derive the corresponding
family of area spectra (\ref{spectab}), which now contain the Casimir
operators of \SL \cite{sergei2,sergei3}.
The main technical difficulty of this method is
the resulting non-commutativity of the \SL connection.
An expression of $\{{\cal A},{\cal A}\}_D$
computed in \cite{sergei4} can be found in appendix \ref{B}.

From this covariant approach, it is easy to
reconstruct the variables of the \SU approach.
Indeed, a suitable projection of the canonical \SL connection
$A_i^X$ ($X$ is an $sl(2,C)$ index) gives an equivalent of
the Ashtekar-Barbero connection:
\be
\iAash{i}{X}
=I_{(Q)Y}^X(\delta^Y_Z-{\im}\Pi^Y_Z) A_i^Z.
\label{AshBar}
\ee
Namely, as is shown in appendix \ref{C}, its three non-vanishing
components $\iAash{i}{a}$
coincide (up to $\im$) with the quantity (\ref{connecbarros})
and thus coincide with the Ashtekar-Barbero connection in the time
gauge $\chi=0$.
Further, they form a connection of the "boosted" subgroup \sgchi,
constructed explicitly in Appendix \ref{C},
when $\chi$ is constant over the canonical hypersurface.
Moreover, despite the fact that
the canonical connection $A$ is non-commutative,
$\Aash$ commutes with itself (see appendix \ref{B}).
Thus, taking $\Aash$ as a canonical variable,
in any gauge $\chi=const$, the phase space has the same structure
as in the usual \SU approach.

This consideration allows to reproduce the phase space of the \SU approach
at the classical level.
However, it simply amounts to break the covariance of the theory and
translate the \SU connection variables into the new formalism:
it is not equivalent to derive the \SU setting
from a covariant quantum theory.
The reason is that $\Aash$ can be considered as an \SU connection,
whereas the covariant loop quantization should be based on a Lorentz
connection. Remarkably, there exists such a Lorentz connection
which is a natural generalization of the Ashtekar-Barbero one \cite{sergei3}.
We describe its properties and the results, which it leads to,
in the next paragraph.

\subsection{The covariant Ashtekar-Barbero connection}

\label{CAB}

The canonical analysis done in preserving covariance,
as described in appendix A, leads to a two-parameter family
of Lorentz connections (\ref{consh}) with such commutation relations
that the corresponding Wilson lines are eigenstates of the area
operator.
As it had been noted in \cite{sergei3}, if we choose the parameters
of the connection, $a$ and $b$ in equation (\ref{consh}),
to be as follows
\begin{equation}
a=-\im, \qquad b=1,
\label{parSU2}
\end{equation}
we obtain a connection, which reproduces the results of the $SU(2)$ approach.
Indeed, it takes form
\be
\SSA_i^X =
I_{(Q)Y}^X(\delta^Y_Z-{\im}\Pi^Y_Z) A_i^Z
- \im \R^{X}_{Y}\ff_i^{Y}(\tQ)
=\iAash{i}{X}-\im \R^{X}_{Y}\ff_i^{Y}(\tQ),
\label{conSU2}
\ee
where $\ff^X_i(\tQ)$ is a function of $\chi$
only:\footnote{Let us note that the same expression arises naturally
when one extends the Ashtekar-Barbero connection by technics
of differential geometry (see appendix \ref{so3toso31}).}
\be
\ff^X_i(\tQ)
=\left(- \frac{\eps^{abc}\chi_b\p_i\chi_c}{1-\chi^2},
\frac{\p_i\chi^a}{1-\chi^2} \right).
\ee
This particular connection possesses the following properties:
\begin{itemize}
\item
For $\chi$ constant on the hypersurface it coincides with $\Aash$
from equation (\ref{AshBar}) and, in particular, for
the ``time gauge'' $\chi=0$, it coincides with
the Ashtekar-Barbero $SU(2)$ connection,
thus being its Lorentz generalization:
\be
\SSA_i^X\mathop{=}\limits_{\chi=0} (0,\frac12 {\eps^a}_{bc}\omega_i^{bc}-
\im\omega_i^{0a}).  \label{conAB}
\ee
\item
As the Ashtekar-Barbero connection, it is commutative (see appendix \ref{B})
\be
\{ \SSA_i^X, \SSA_j^Y\}_D = 0.
\ee
\item
Its commutator with the triad multiplet is (see equation (\ref{newAiP}))
\be
\{ \SSA_i^X,\tQ_Y^j\}_D=-\im \delta_i^j I_{(Q)Y}^X,
\ee
where $I_{(Q)}$ is the projector on the $SU(2)_{\chi}$
part of the Lorentz group.
\end{itemize}

Due to this last relation,
the area spectrum given by this $SL(2,C)$ connection
coincides exactly with the one coming from Loop Quantum Gravity
given by the Casimir operator of $SU(2)$:
\be
{\cal S}\sim \hbar \im\sqrt{ C(su(2))}.
\label{asSU2}
\ee

\subsection{SU(2) Hilbert space from covariant quantization}
\label{su2} \label{Hil}

Having in hands a Lorentz connection reproducing classically
the properties of the \SU Ashtekar-Barbero connection,
we can ask if this relation is maintained at the quantum level.
It turns out that choosing the connection (\ref{conSU2})
as a basic variable, the covariant theory can be relatively easy
quantized. In particular, the second class constraints can be properly
imposed at the quantum level.
Then, in a particular gauge, we actually recover the usual
\SU spin network Hilbert space of LQG.
Thus, we are able to derive the \SU LQG from the quantized
covariant formulation.

Having a Lorentz connection,
the natural objects to
consider are the holonomies or the Wilson lines
\be
U_{\alpha}[\SSA]={\cal P}\exp\left(\int_\alpha dx^i \SSA_i^X T_X\right),
\label{wl}
\ee
where $\alpha$ is an oriented path and $T_X$ are the \SL generators.
Taking their trace
for a closed loop $\alpha$ in a given representation of $SL(2,C)$,
we obtain gauge invariant objects.
We usually look at the
representations $R^{(n,\rho)}$ from the principal series of
unitary irreducible representations of \SL (see appendix \ref{D}
for details), since they are the ones entering the Plancherel
formula.
Such observables can be generalized to arbitrary oriented graph
and give rise to spin networks \cite{spinnet}. However, this
construction is not
enough in our case because these functionals are not eigenvectors of the
area operator and do not have a direct physical interpretation.
Indeed, at a given point $x$ of intersection of the
loop and a small surface whose area we are computing, we need to
decompose the representation $R^{(n,\rho)}$ of \SL into
representations $V^j_{\chi(x)}$ of the subgroup \sgi{\chi(x)},
which leaves the vector $\chi(x)$ invariant. According to equation (\ref{asSU2}),
each subspace $V^j$ will contribute $\im\sqrt{j(j+1)}$
and the overall area operator will not be simple multiplication on the
\SL Wilson line.

In order to get an eigenvector, we need to select a particular
subspace $V^j_{\chi(x)}$. Since this subspace depends on the field $\chi$,
this leads us to consider gauge invariant functionals of
both the connection $\SSA$ and the time normal field $\chi$.
Notice, that it is consistent due to the relation (\ref{cAchi}).
Gauge invariance will then read
\be
f(\SSA,\chi)=f(^g\SSA=g\SSA g^{-1}+g\partial g^{-1},
{}^g\chi=g\cdot \chi).
\ee
Such invariant functions are in fact entirely given by the functions
$f_{\chi_0}(\SSA)=f(\SSA,\chi={\chi_0})$ taken for $\chi$ constant on the
hypersurface equal to ${\chi_0}$.
The remaining gauge symmetry of $f_{\chi_0}$ is only an \sg
gauge symmetry and is compact. The choice of section
"$\chi=\chi_{0}$" will be called the time gauge.

Following the ideas of loop quantum gravity where one considers cylindrical
functions of the Ashtekar-Barbero connection,
we introduce cylindrical functions $f_{\Gamma}(\SSA,\chi)$ which will be
constructed on an oriented graph $\Gamma$. They will depend only on
the holonomies $U_1,\dots,U_E$ of $\SSA$ along the edges of $\Gamma$
and on the values $\chi_1,\dots,\chi_V$ of the field $\chi$
at the vertices of $\Gamma$. The gauge invariance will
then read\footnote{In fact, the $\chi$ field is a
vector field $(1,\chi^a)$, with $\chi ^2 \le 1$.
One can normalize the
time normal so it is represented by a vector living on the (upper)
hyperboloid of the Minkovski space, as in the spin foam context
\cite{psn}. This defines a vector field
$$
x=\left(\f{1}{\sqrt{1-\chi^2}},\f{\chi^a}{\sqrt{1-\chi^2}}\right).
$$
Then, the transformation law of $x$ is simply the usual Lorentz
transformation in the Minkovski space. This is what is implicit in the
definition of the new cylindrical functions and the projected spin
networks. Moreover, using this new field can simplify expressions of
some functions of $\chi$ such as
$$
I_{(P)X}^Y=
\left(
\begin{array}{cc}
{\delta_a^b x_0^2-x_ax^b} &
{\eps_a}^{bc}x_0x_c \\
{\eps_a}^{bc}x_0x_c &
-(\delta_a^b{\vec x}^2-x_ax^b)
\end{array} \right).
$$}
\be
f_{\Gamma}(U_1,\dots,U_E,\chi_1,\dots,\chi_V)=
f_{\Gamma}(g_{t(e)}U_eg_{s(e)}^{-1},g_v\chi_v)
\textrm{ with } g_v \in SL(2,C),
\ee
where $s(e)$ ($t(e)$) is the source (target) vertex of edge $e$.
One can use the Haar measure on $[SL(2,C)]^E$ to introduce a
(kinematical) scalar product on the space of $L^2$ gauge invariant
functions:
\be
\langle f | g\rangle=
\int_{[SL(2,C)]^E} dU_e \,\overline{f_{\Gamma}(U_e,\chi_v)}
g_{\Gamma}(U_e,\chi_v).
\label{kineprod}
\ee
This scalar product does not depend on the choice of
$(\chi_1,\dots,\chi_V)$ due to the Lorentz invariance.
We denote ${\cal H}_0$ the resulting Hilbert space.
Let us emphasize that this will not be the physical Hilbert space
since it is likely that we will need to modify the scalar product
\Ref{kineprod} to take into account the second class constraints.
Nevertheless, exhibiting a basis of ${\cal H}_0$ sheds light
on the structure of the theory.

To construct it, we take the usual \SL spin networks, and insert
a projector $\Ppr{j}{\chi_v}:R^{(n,\rho)}\rightarrow V^j_{\chi_v}$
at each edge around every vertex $v$.
This procedure is equivalent to the change of the Wilson lines \Ref{wl}
by Wilson lines {\it projected at the ends}
\be
\CU^{(j_{s(e)},j_{t(e)})}_{e}[\SSA,\chi]=
\Ppr{j_{t(e)}}{\chi_{t(e)}} U_{e}[\SSA]
\Ppr{j_{s(e)}}{\chi_{s(e)}}.
\label{WLpe}
\ee
The projector can be written as
\be
\Ppr{j}{\chi}=(2j+1)\int_{SU_{\chi}(2)}dh\, \overline{{C}^j(h)}D(h) ,
\ee
where $C^j$ is the character of the \sg representation $j$ and
$D(h)$ is the representation matrix of the group element $h$.
It is important that the projector depends on
$\chi$. Due to this dependence it transforms homogeneously under
Lorentz boosts
\be
\Ppr{j}{{}^g\chi}=D(g)\Ppr{j}{\chi} D^{-1}(g),
\label{tranpr}
\ee
as well as the Wilson lines \Ref{WLpe}.
Therefore, the resulting {\it projected spin
networks} are still gauge invariant and belong to our Hilbert space
${\cal H}_0$.
Due to the projections, they are labelled by
one \SL representation $(n_e,\rho_e)$ for each edge $e$, two
\sg representations for each edge $(j_{s(e)},j_{t(e)})$
(for the source and target vertices of $e$)
and \sg intertwiners at all vertices \cite{psn}.
It is straightforward to check that two projected spin networks
with different labels are orthogonal with respect to the scalar product
\Ref{kineprod}.
Their completeness is also evident.
Thus, the projected spin networks realize an orthonormal basis
in ${\cal H}_0$.

Moreover, such states are eigenvectors of the area operators
of surfaces intersecting the spin networks at vertices,
and the area attached to one edge at a vertex is given by the
\sg representation $j$ attached to the corresponding end of the edge:
\be
{\cal S}\sim \im\sqrt{j(j+1)}.
\ee
Interestingly, this does not depend at all on the \SL
representations. What are they here for?
For the moment, they give the way the projected
spin networks change under \SL gauge transformations.
We can say that the \sg representations define the space
geometry while the \SL representations give its space-time
embedding and define how it gets modified under boosts.
However, we have not finished the job yet and we still need to take
into account the second class constraints.

The second class constraints now correspond to
the constraints satisfied by the connection \Ref{conSU2}:
\be
I^X_{(P)Y}\SSA_i^Y=\Pi^X_Y \ff_i^Y(\tQ). \label{conA}
\ee
Through this relation, $\SSA$ depends explicitly on $\chi$, and
this reduces the number of independent components of $\SSA$
from 18 to 9.
The physical meaning of the constraints becomes obvious in the time gauge,
when one rotates $\chi$ to $\chi_0$ on all the hypersurface.
Then we have $I_{(P)}\SSA=0$.
As a result $\SSA$ is reduced to its \sgi{\chi_0} part
and becomes simply an \SU connection.
Computing the holonomies of $\SSA$, we get group elements
belonging to the \sgi{\chi_0} subgroup.
This has an immediate consequence that the projected Wilson lines \Ref{WLpe}
are non vanishing only for $j_{s(e)}=j_{t(e)}$ and produce the usual \SU Wilson lines:
\be
\CU^{(j_{s(e)},j_{t(e)})}_{e}[\SSA,\chi_0]=\delta_{j_{s(e)}j_{t(e)}}
\iota\left(U_{e}[\Aash]\right),
\label{WLpesu}
\ee
where $\iota$ denotes the embedding of an \SU group element into a
representation $R^{(n,\rho)}$ of $SL(2,C)$.
In fact, since the result does not depend on
the \SL representation, we can omit this embedding provided it was chosen
so that $j\ge n$. Otherwise the representation $j$ does not enter
the decomposition of $R^{(n,\rho)}$ over the subgroup and the projection
(\ref{WLpe}) vanishes. Therefore, it is enough to restrict ourselves
from the very beginning to one arbitrary simple representation of type
$R^{(0,\rho)}$ since each of them contains in its decomposition all
spectrum of \SU representations.

Thus, from \Ref{WLpesu} we obtain that each edge is labelled by only
one \sg representation $j_e$ and associated with an \SU group element.
Then, the projected spin networks, evaluated in the time gauge,
reduce to the usual \SU spin networks, which are actually the natural
\SU gauge invariant objects.
The scalar product, which takes into account the reduction of
the configuration space induced by the second class constraints,
is not anymore the kinematical one but
\be
\langle f | g\rangle=
\int_{[SU_{\chi_0}(2)]^E} dU_e \,\overline{f_{\chi_0}(U_e)}
g_{\chi_0}(U_e).
\label{su2prod}
\ee
We have actually recovered the full (kinematical) structure
of \SU Loop Quantum Gravity at the level of the Hilbert space.

Up to now, we have described how covariant functions of the
connection and the time normal field, which are solutions to the second
class constraints, look like in the time gauge.
However, we would like to be more ambitious and describe
the physical Hilbert space out of the time gauge, {\it i.e.},
characterize the space of
all gauge invariant functionals of Lorentz connection and $\chi$,
which are nontrivial only for the solutions of the constraints.
This will complete the implementation
of the second class constraints at the quantum level.


Let us have a closer look to the situation in the time gauge.
One can note that
for the Lorentz connection satisfying \Ref{conA} the insertion of
the projector on a representation $j$ in the middle of
an edge has a trivial effect: if $j_{s(e)}=j_{t(e)}=j$ we get identity,
otherwise the result vanishes.
Therefore, we can infinitely refine each edge of the initial graph
by adding infinite number of bivalent vertices. Each of them introduces
the corresponding projector so that the refinement is equivalent to
consider the following {\it fully projected Wilson lines} \cite{sergei4}:
\be
U^{(j)}_{\alpha}[\SSA,\chi]= \lim\limits_{N \rightarrow \infty}
{\cal P}\left\{ \prod\limits_{n=1}^{N}
\Ppr{j}{\chi_{_{v_{n+1}}}} U_{\alpha_n}[\SSA] \Ppr{j}{\chi_{_{v_{n}}}} \right\},
\label{WLp}
\ee
where $\alpha=\bigcup_{n=1}^N \alpha_n$ is
a partition of the line into small pieces.
As we just showed, this procedure does not
change the projected spin network for the connection $\SSA$:
\be
\CU^{(j,j)}_{\alpha}[\SSA,\chi]=U^{(j)}_{\alpha}[\SSA,\chi].
\ee

Now we prove that provided each edge is associated
with a simple representation $R^{(0,\rho)}$ of $SL(2,C)$,
such infinitely refined projected spin networks
depend only on the solution of the second class constraints \Ref{conA}.
This statement follows from the property of the Lorentz generators:
\be
\Ppr{j}{\chi}F_a \Ppr{j}{\chi}=\im_{(j)} \Ppr{j}{\chi}H_a \Ppr{j}{\chi},
\qquad \im_{(j)}=\frac{n\rho}{j(j+1)},
\label{prop}
\ee
where $H_a$ are generators of the \sgchi subgroup,
$F_a$ are the corresponding boost generators.
For the simple representations $\im_{(j)}=0$ which
implies that the projected boost generators then vanish.
On the other hand, the infinite refinement \Ref{WLp} is equivalent to such
projection in the exponent of Wilson lines (see \cite{sergei4}),
since for sufficiently fine partition, we can write:
\be
U^{(j)}_{\alpha}[A,\chi]= {\cal P}\left\{ \prod\limits_n
\Ppr{j}{\chi_{_{v_{n+1}}}}
\left(1+\int_{\alpha_n}dx^i A_i^X T_X\right) \Ppr{j}{\chi_{_{v_{n}}}}
\right\}
\label{WLp-su2}
\ee
and the projectors act directly on the connection $A_i^X T_X$.
Therefore, in the time gauge it has an effect to change a projected Wilson
line \Ref{WLpe} with $j_{s(e)}=j_{t(e)}=j$
by an \SU Wilson line in the representation
$j$ dependent only of \sgi{\chi_0} components of the connection.
This is just the same result as the connection $\SSA$ gives.
To leave the time gauge it is enough to make a gauge transformation.
Due to the Lorentz invariance the infinitely refined projected spin networks
still give a solution of our problem representing a basis of the
physical Hilbert space.

It is important to note that the constructed states are eigenvectors
of all area operators \cite{sergei4}.
This result is provided by the infinite refinement:
due to this, each point of an edge can be considered as a vertex.
Thus, our states possess all properties of the \SU spin
networks and are their Lorentz generalization.


\subsection{Drawbacks of the $SU(2)$ formalism}

Thus the SU(2) LQG can be rigorously derived from the quantization of
the covariant formalism based on the connection (\ref{conSU2}).
This allows to look at its status from the point of view of
the covariant quantization.
First of all, let us elucidate which problems the SU(2) LQG possesses
and whether they can be solved in the covariant formalism.

Apart from the problem of implementing the right Hamiltonian operator,
there are two main issues. The first one is the Immirzi parameter $\im$
\cite{imm}. It appears in the classical theory  parametrizing
different Ashtekar--Barbero connections of the \SU approach,
which all are related by a canonical transformation.
However, this parameter enters the area spectrum.
Therefore, that canonical transformation
can not be implemented by a unitary operator and the resulting
quantum theories are inequivalent.
We discuss the physical relevance of the Immirzi
parameter in section \ref{section:immirzi}.
Nevertheless, let us note that
there does not exist any canonical transformation, relating
theories with different values of $\im$, within our covariant formalism.

The next problem is the loss of the spacetime interpretation for the SU(2)
connection \cite{Sam}. Although it has not been taken into account
seriously, it has deep consequences. In particular, this fact is probably
the reason why the quantum constraint algebra
does not reproduce the classical one and contains an anomaly
\cite{qca,BS,RR}.

It turns out that the covariant formalism is very convenient to address
this second problem. In terms of transformation properties it means that
the Ashtekar--Barbero connection cannot be extended in such a way that
it transforms as a true space-time connection under
4-dimensional diffeomorphisms.
And indeed, it was shown that its Lorentz extension (\ref{conSU2})
does not tranform correctly under the time diffeomorphisms
\cite{sergei3}.
As a result, the implementation of this symmetry in quantum theory
in the framework of loop quantization fails. The reason is that
loop operators are not mapped to the time translated loop operators
after symmetry transformation.\footnote{The situation is essentially
the same as it would be if the connection does not transform correctly
under the space diffeomorphisms. Then there would not be easy way to realize
this symmetry on the space of loop states.}
Hence the quantum diffeomorphism algebra contains an anomaly.

However, may it not be a problem but an unavoidable property
of quantum gravity? The answer depends on whether one can find a
quantization
preserving the full diffeomorphism invariance. If such a quantization
exists, of course, it should be considered as more preferable,
since the whole history of quantum theory tells that one should try to
preserve classical symmetries as much as possible, especially,
when they are as fundamental as the diffeomorphism invariance is
believed to be.

To answer this question, let us recall that the covariant
Ashtekar--Barbero connection (\ref{conSU2}) is only one among
the two-parameter family of Lorentz connections found in the covariant
approach. In principle, each of the connections could be used in
loop quantization and each would lead to different physics (for example,
different area spectra).
Thus, they represent a real quantization ambiguity of
the loop approach. Could this ambiguity be resolved?
Is it possible to find a criterion which allows to choose the right
connection?

The answer is affirmative and the corresponding criterion is actually
simply that it transforms properly under
the time diffeomorphisms.
Indeed, it turns out that if we impose this additional restriction,
there is only one connection satisfying it \cite{sergei3},
{\it i.e.}, possessing a genuine spacetime interpretation.
This means that there is a {\it unique} loop quantization preserving all
classical symmetries of general relativity.
Moreover, for this choice of connection the area spectrum does not depend
on the Immirzi parameter. This gives an additional evidence in favour of
such quantization and shows that all the problems appearing in SU(2)
LQG are likely to find their
solutions in the covariant approach.

\section{Quantization preserving diffeomorphism invariance}

\subsection{Canonical structure and area spectrum}

In this section we describe the unique spacetime Lorentz connection
diagonalizing the area operator and the resulting quantum picture.
This connection corresponds to the choice $a=b=0$ in equation (\ref{consh})
what leads to the following shifted connection \cite{sergei2,sergei3}
\be
\SA_i^X=A_i^X + \frac{1}{2\left(1+\frac{1}{\im^2}\right)}
\R^{X}_{S}I_{(Q)}^{ST}\R_T^Z f^Y_{ZW}\Pt_i^W \G_Y.  \label{spcon}
\ee
In this case the Dirac brackets can be given in the simple form:
\beq
\{ \tP_X^i,\tP_Y^j\}_D&=&0, \\
\{ \SA_i^X,\tP_Y^j\}_D&=&\delta_i^j I_{(P)Y}^X ,
\label{comm}
\eeq
whereas the commutator of two connections is much more
complicated (see appendix \Ref{B}) except from the relation:
\be
\{I_{(P)}R^{-1}\SA,I_{(P)}R^{-1}\SA\}_D=0.
\ee
The area operator following from the commutation relations \Ref{comm}
is expressed as a combination of two Casimir operators:
\be
{\cal S}\sim \hbar \sqrt{C(su(2)) - C_1(so(3,1))}.
\label{spectr}
\ee

As the connection \Ref{conSU2}, $\SA$ satisfies some constraints reducing
the number of its independent components. There are 3 such constraints
\cite{sergei2}. However, we can use additional ambiguity
to add to any quantity a combination of the second class constraints
\Ref{psi} in order to remove 6 more components,
without modifying any commutation relations.
The most natural choice is
\beq
\tSA_i^X &=&\SA_i^X- \frac12 R^X_{Y}
\left(\Qt_l^Y(\Qt\Qt)_{ik}-\frac12 \Qt_i^Y(\Qt\Qt)_{lk}\right)\psi^{lk}
\nonumber \\
&=& \left(1+\frac{1}{\im^2}\right) I_{(P)Y}^X (R^{-1})^Y_Z A_i^Z
+ R^X_Y \Gamma_i^Y,
 \label{spconnew}
\eeq
where
\be
\Gamma_i^X = \frac12 f^{W}_{YZ}I_{(Q)}^{XY} \Qt_i^Z \p_l \tQ^l_W
+\frac12 f^{ZW}_Y\left((\Qt\Qt)_{ij} I_{(Q)}^{XY}
+\Qt_j^X\Qt_i^Y -\Qt_i^X\Qt_j^Y \right) \tQ^l_Z \p_l \tQ^j_W.
\label{gam}
\ee
It is clear then that the connection \Ref{spconnew} satisfies the constraints:
\be
I_{(Q)Y}^X\tSA_i^Y=\Gamma_i^X(\tQ).
\label{contA}
\ee

Let us note the most important differences in comparison with
the case of the Lorentz generalization of the Ashtekar-Barbero connection
described in section \ref{CAB}.
\begin{itemize}
\item
The nontrivial part of the connection is contained in the boost rather
than \SU components (see \Ref{comm} and \Ref{contA}).
\item
The nondynamical part of the connection given by $\Gamma_i^X(\tQ)$
does not vanish in the time gauge. It gives actually a generalization of
the Christoffel connection and it is defined by the triad field.
\item
The connection remains noncommutative.
\item
The commutation relations \Ref{comm} and the area spectrum
does not depend on the Immirzi parameter.
\end{itemize}

All these differences have deep consequences for quantization.
First of all, the noncommutativity of the connection brings doubts
about the validity of the connection representation and the use of
loop functionals as configuration variables.
Nevertheless, we can try to ignore this difficulty for a moment
with a hope to resolve it in the end of the way.\footnote{An example of such
sitation, when we end up with commutative spin networks, can be found in
\cite{sergei4}. Another way would be to look for a triad representation.}
Then we try to carry out the same program which was realized in
section \ref{Hil}. And we do not encounter any problems in the first part.
The construction of the projected spin networks based on
Wilson lines \Ref{WLpe} does not refer to particular properties of
the used connection and it is still valid for any Lorentz connection.
In this way we end up with the same kinematical Hilbert space ${\cal H}_0$.

However, we are not able to carry out the second part of the program and
solve the second class constraints
on this Hilbert space as we had done for the connection $\SSA$.
Indeed, we should somehow take into account at the quantum level that we
fix the \SU components of the connection.
Moreover, this fixed value should be determined by the triad $\tE$,
which is difficult to realize using only functionals of connection. Maybe,
this problem can be solved in a triad representation as done with the
reality conditions for the self-dual (complex) Ashtekar formulation
corresponding to the case $\im=i$ \cite{triadrep}.
This will be investigated in future work.
Thus, the problem of solution of the second class constraints for the
spacetime connection at the level of Hilbert space looks quite nontrivial
and remains to be done.

Nevertheless, it is possible to sidestep this problem and impose the
second class constraints at a finite number of points. Indeed,
as explained in the next paragraph, it
turns out that projected spin networks projected onto the trivial
$SU(2)$ representation $j=0$ solve the second class constraints at
their vertices and give the same Hilbert space of quantum states
as obtained from the spin foam approach.

\subsection{Recovering the spin foam basis}

Spin foam models are the space-time models corresponding to the
evolving spin networks from LQG.
Up to now, there has been lacking an explicit link between the
existing spin foam models,
which are based on $SL(2,C)$, and the canonical framework, which is based
on the $SU(2)$ symmetry group.
The present covariant canonical framework
builds a bridge between these two pictures and this may help to build
a consistent quantum space-time picture.

The most promising spin foam model for both Euclidean
and Lorentzian gravity is the Barrett-Crane model \cite{bc1,bc2}.
Its construction relies on methods from geometrical quantization \cite{bc1,bb}, but
can also be related to the generalised Palatini action (with Immirzi parameter)
through a generalised constrained BF theory \cite{bf,etera1,etera2}.
From the point of view of canonical quantization, it can be interpreted as
quantizing the system
without imposing the second class constraints and imposing them only
afterwards at the quantum level. They can then be translated into
the so-called simplicity constraints restricting the used $SL(2,C)$
representations to simple ones, which have a vanishing quadratic Casimir
operator \cite{freidel,etera1,etera2}.

More precisely, when carrying the canonical analysis \cite{sergei1},
we get second class constraints of two types
$\phi^{ij}=\Pi^{XY}\tl{Q}^i_X\tl{Q}^j_Y$ \Ref{phi}
and $\psi^{ij}$ (see \Ref{psi} for an explicit expression).
These constraints give raise to a Dirac bracket. Now, an interesting
property of this Dirac bracket is that it is equal to the initial
Poisson bracket when both the considered quantities commute with only the
$\phi^{ij}$ constraints:
\be
\{K,\phi\}=\{L,\phi\}=0
\Rightarrow
\{K,L\}_D=\{K,L\}.
\ee
This leads to think that considering only quantities that commute
with the $\phi$ constraints allows to ignore the $\psi$ constraints.
The advantage of such a viewpoint is that the $\phi$ constraints
seem much easier to implement than the $\psi$ constraints.
Moreover, it coincides exactly with the simplicity constraint
used in spin foam models. More precisely, let us consider a spin network
based on the initial $SL(2,C)$ connection $A$ and pick
a point on a given link of the graph.
Imposing that it commutes with $\phi$ leads to an equation on
the Casimir operators of the representation living on the chosen link:
\be
\f{2}{\beta}C_1(SL(2,C))=(1-\f{1}{\beta^2})C_2(SL(2,C)),
\label{wrong}
\ee
where $C_1=g^{XY}T_XT_Y$ and $C_2=\Pi^{XY}T_XT_Y$ are
the two Casimirs of $SO(3,1)$. This equation is exactly the same
as the one arising in the construction of spin foam models
from the generalised Palatini action \cite{etera1}.
Within the spin foam context, it was argued
that such an equation is meaningless and it turned out that there
exists an ambiguity in the quantization procedure which allows
to rotate the constraint to the usual simplicity one $C_2=0$
\cite{etera2}.
This ambiguity is to be compared to the
two-parameter ambiguity $\SA(a,b)$
in the choice of a $SL(2,C)$ connection
in the covariant canonical frame \Ref{consh}:
does taking full account of the second class constraints through
the Dirac bracket and choosing the space-time connection $\SA=\SA(0,0)$
cancel the rotation introduced by $\beta$ and lead to the same result
obtained for spin foams?

\medskip
To start with, let us look at the case $\beta\rightarrow\infty$. This
corresponds to the usual Palatini action without the extra term
introduced by the Immirzi parameter, and we have the following Poisson
bracket relations:
\beq
\{I_{(P)}\SA,I_{(P)}\SA\}_D &=&0, \nonumber \\
\{\tP,\tP\}_D &=& 0, \nonumber \\
\{ \SA_i^X,\tP_Y^j\}_D&=&\delta_i^j I_{(P)Y}^X.
\label{poissonrelation}
\eeq
We would like to ignore $I_{(Q)}\SA$
(which does not come in the commutation relation with the $P$ field,
and could be completely constructed as an operator from the $P$ operator
at the quantum level)
and construct functionals
depending only  on $I_{(P)}\SA$. To this purpose, we consider simple
spin networks, which are the $j=0$ case of the projected spin networks
introduced in \ref{su2} \cite{psn}. We choose a fixed (oriented)
graph, whose edges we label with $\sl$ representations $(n_e,\rho_e)$.
We construct the holonomies $U_e$ of the connection $\SA$
along these edges. We consider their trace on the $SU(2)$ invariant
subspace of the edge representation.
This is the $j=0$ subspace, and its existence
selects out the simple representation $(n_e=0,\rho_e)$
which we will
note simply $\rho_e$. Finally, the simple spin network functionals are:
\be
s^{\{\rho_e\}}(U_e)=
\prod_e \langle\rho_e\,\chi_{s(e)}\,j=0|U_e
|\rho_e\,\chi_{t(e)}\,j=0\rangle,
\ee
where $\chi_v$ is the value of the field $\chi$ at the vertex $v$ and
$|\rho\chi_vj=0\rangle$ is the vector in the $\rho$ representation which
is invariant under $SU(2)_{\chi_v}$.
These are cylindrical functionals of the connection $\SA$ and depend
only on $I_{(P)}\SA$ at the vertices. Therefore, the Poisson bracket
of two such functionals whose graphs intersect only at some common
vertices vanishes. This is understandable within the spin foam context,
where we have a complete discrete view of space-time. Only the
vertices are (space-time) points, then an edge is a relation between
two points and is not considered as a continuous line of points.
Indeed only at the vertices, we do know the time normal. From this
point of view, considering two simple spin networks, if their graphs
intersect, then the intersection point is to be defined, and so it
should be a vertex of the two graphs.
Thus, either it is already a common vertex, or
we should refine the graph (add a bivalent vertex in the middle of the
edges) so that it becomes one.
Moreover, we can compute the action of the area operator
of a surface intersecting the graph at the end of an edge (at the
vertex). The simple spin network is its eigenvector with
eigenvalue:
\be
\CS\sim\sqrt{-C_1(\sl)}=\sqrt{\rho_e^2+1}.
\ee
This spectrum is always well-defined,
corresponds truly to a space-like surface
and is, in fact, exactly the same area spectrum as obtained through the
spin foam approach.
To sum up, we can choose an initial set of points on the manifold
which will be the vertices of all the considered graphs,
then we consider the simple spin networks based on such graphs and we
obtain a representation of the initial commutation relations and
quantum structure which reproduces exactly the kinematical setting
and the boundary states of the Barrett-Crane model \cite{psn}.
This representation takes into account only the commutation relations
\Ref{poissonrelation} at the chosen points.
In principle,
 we could build a delicate ladder of operators taking into account
the necessity of considering the commutator of two simple spin
networks only when their graphs intersect at common points, then it
would be possible to
get a complete representation of the Poisson brackets
\Ref{poissonrelation}.

This shows that the second class constraints build a theory based on
the coset $SO(3,1)/SO(3)$ or $SL(2,C)/SU(2)$ just as in the spin foam
scheme. This result opens the door to a generalisation to higher
dimensions: it seems possible to reproduce, within the canonical
approach, the spin foam result that simplicity conditions impose a
$SO(D)/SO(D-1)$ coset structure to the quantum theory \cite{coset}.

\medskip
Now let us look at the general case with $\beta$ arbitrary. The Poisson
algebra reads:
\beq
\{I_{(P)}R^{-1}\SA,I_{(P)}R^{-1}\SA\}_D &=&0, \label{new1} \\
\{\tP,\tP\}_D &=& 0,  \label{new2}\\
\{ \SA_i^X,\tP_Y^j\}_D&=&\delta_i^j I_{(P)Y}^X.
\label{new3}
\eeq
The situation is complicated by the $R$ change of basis.
And it is not obvious how the above quantization procedure
generalises in this case. Indeed the commutation relations between
$\SA$ and $\tP$ invite us to consider the same functionals as
previously. Nevertheless, the commutation relations $\SA$ and itself
tell us that the operator for $I_{(P)}\SA$ will then not be trivial.
The solution to the problem is to use the (second class) constraint
$I_{(Q)Y}^X\tSA_i^Y=\Gamma_i^X(\tQ)$.

First, we can change $\SA$ to $\tSA$
without modifying anything in the Poisson brackets. Then, as
$I_{(Q)}\tSA$ commutes with $\tP$, we get
\be
\{I_{(P)}R^{-1}\tSA,\tP\}_D=
\{I_{(P)}\tSA-\f{1}{\beta}\Pi I_{(Q)}\tSA,\tP\}_D=I_{(P)}.
\label{new4}
\ee
Then, one has the same structure as previously
replacing $I_{(P)}\SA$ of the case $\beta\rightarrow\infty$
by $I_{(P)}R^{-1}\tSA$. One considers simple spin networks
constructed with the connection $\tSA$. The operator
$\widehat{\tilde{P}}$
is still the derivation with respect to $\tSA$ and its action is
the insertion of the generators $T$ of \SL. The operator
$\widehat{I_{(Q)}\tSA}$
can be deduced as the Christofel symbol $\Gamma(\tilde{Q})$
constructed with the $\widehat{\tilde{Q}}$ operator.
Then, we can choose the operator
$\widehat{I_{(P)}R^{-1}\tSA}$ to be the multiplication by $I_{(P)}\tSA$.
This operator commutes with itself as any multiplication,
which realises \Ref{new1}. Moreover,
the commutator of the multiplication $I_{(P)}\tSA\times$ with the derivation
operator $\widehat{\tilde{P}}$ gives the identity, so that this choice
of quantizaion realises the Dirac bracket \Ref{new4} and thus the
bracket \Ref{new3}. Therefore, we have a complete realisation of
the Dirac brackets of the connection and the triad. Finally, we can deduce
the operator for $I_{(P)}\tSA$ from the operator for $I_{(P)}R^{-1}\tSA$:
\be
\widehat{I_{(P)}R^{-1}\tSA}=\widehat{I_{(P)}\tSA}-
\f{1}{\beta}\Pi\widehat{I_{(Q)}\tSA}
\Rightarrow
\widehat{I_{(P)}\tSA}=
I_{(P)}\tSA\times +\f{1}{\beta}\Pi\Gamma(\widehat{\tilde{Q}}).
\ee
This concludes the spin foam quantization in the canonical framework and shows
that the Immirzi parameter $\beta$ is not an obstacle to the quantization
as had suggested the modified simplicity condition \Ref{wrong}.

\section{Physical Relevance of the Immirzi parameter}

\label{section:immirzi}

From the very beginning we introduced the parameter $\beta$
in the theory. It was identified as the Immirzi parameter
of \SU LQG through the canonical analysis in the time gauge \cite{holst}.
At the classical level it does not change the equations
of motion and, therefore, it does not influence the dynamics.
Does it play any role at the quantum level?
The SU(2) LQG says it does. The main reason is that the physical
spectra of geometrical operators, like the area spectrum (\ref{asSU2}),
depend on it. Besides, it comes in the Hamiltonian constraint and thus
modifies a priori the dynamics of the theory.
As a result, the Immirzi parameter should become physical
and it should be considered as a new
fundamental constant \cite{rt}. It is usually argued that it can be
fixed by looking at the black hole entropy.

But we argued that the SU(2) approach is based on a wrong choice of
the connection and there is another choice which seems to be the
only correct one. Does the Immirzi parameter become physical
in this second quantization?
The results of section III say it does not.
The main reason is that it does not appear in the commutation relations
of the connection and triad mutliplet (\ref{comm}). It is this commutator
from which we derive the spectrum of area and other geometrical
operators. Of course, the Immirzi parameter appears in the commutator of
two connections (\ref{AAcom}), but only in the universal form
in the prefactors. Therefore, even if this commutator contributes to some
physical results, it is very likely that the Immirzi parameter will be
cancelled.

One more evidence that the Immirzi parameter remains unphysical is given
by the path integral quantization. It was shown that the formal path integral
constructed for the generalized Hilbert-Palatini action does not depend
on $\beta$ \cite{sergei1}. Note that the path integral does not refer
to any choice of connection but relies only on the BRST analysis
based on the classical symmetry algebra. Similarly, the
spin foam quantization, which can be understood as a discrete path integral,
of the generalized Hilbert-Palatini action was shown not to depend on
the Immirzi parameter \cite{etera1,etera2}. These two results point to
the non-relevance of the Immirzi parameter in the space-time dynamics.

All this allows to conclude that the appearence of the Immirzi
parameter in the results of the SU(2) LQG is a consequence
of the quantum anomaly in the four-dimensional diffeomorphism invariance.
Instead, once the quantization preserving this symmetry is chosen,
it remains unphysical as it was at the classical level.

Moreover, the covariant formalism reveals that
actually there are two ambiguities in the theory: choice of $\beta$
due to the additional term in the generalized Palatini action,
and choice of connection (the parameters $(a,b)$) which is used in
the definition of loop variables.
(This is also true in
the spin foam context \cite{etera1,etera2}, where we have an ambiguity
at the level of the $BF$ action and an ambiguity when quantizing bivectors.)
The first ambiguity is a classical one, whereas the latter is trully
quantum. It is explicily seen from the commutation
relations (\ref{newAiP}) and the area spectrum (\ref{spectab})
found for arbitrary connection of the family (\ref{consh}).
They depend only on the peremeters $a$ and $b$ from the definition of
the connection but not on $\beta$.
The dependence on the Immirzi parameter in the SU(2) case
appears only after the identification $(a,b)=(-\beta,1)$
(\ref{parSU2}).\footnote{In fact, we can obtain the \SU area spectrum
even in the limit $\im\rightarrow\infty$. In this limit
the covariant formalism still exists and possesses the same $(a,b)$
ambiguity in the choice of connection. Choosing $b=1$, we get
the spectrum \Ref{asSU2} given by the Casimir operator of \SU with
$\im$ replaced by $a$. The only diffirence of this formulation
with the previous one is that the connection remains non-commutative.
We need to have a finite $\im$ to get commutativity. From this point of view,
the introduction of $\beta$ and the extra term in the Palatini action
appears like a regularisation.}.
In any case, this quantum ambiguity is fixed by the requirement
to retain the classical symmetry.

\section{Conclusions \& Outlooks}

 In the present work, we have derived the framework of \SU Loop
Quantum Gravity from the explicitly Lorentz covariant formalism.
It was done not only at the classical level, but
quantizing the covariant theory, so that we were able to
reproduce the Hilbert space structure of \SU LQG from
the Hilbert space of covariant quantization.
This was possible due to the existance of a Lorentz generalization
of the Ashtekar-Barbero connection. Choosing this connection
as a basic quantum variable, the quantization
program can be easily carried out. In particular, one can accomlish
the most important step: to implement the second class constraints
at the Hilbert space level.

This derivation allows a new viewpoint on the dynamics of LQG.
Indeed, the covariant approach makes easy to study the space-time symmetries
and whether they are preserved or not through the quantization
process. Because the covariant Ashtekar-Barbero
connection does not transform correctly under time
diffeomorphisms, it is not a space-time connection.
Therefore, there seems to be
a prefered frame defined by the time gauge and the theory seems to
break diffeomorphism invariance.

We have also underlined the existence of a true space-time connection,
the unique Lorentz connection which transforms properly under
space-time diffeomorphisms. We described its properties,
how to quantize the theory with this connection, and the problems
which appear on this way.
In particular, we suggested a quantization procedure which leads
to the resulting picture of quantum geometry states similar to one of
a particular spin foam model --- the Barrett-Crane model. This hints
toward a link between the space-time formalism given by spin foams and
the canonical frame given by LQG and spin network states. One may
hope to better uderstand the geometry defined by spin foams and find the
right dynamics of the theory.

Therefore, it would be very interesting to study the implementation of
the Hamiltonian constraint within this covariant theory, an issue
which we will investigate in future work. Nevertheless, this quantization
procedure linking the canonical theory to the spin foam setting is based on
solutions at a finite number of points to the second class constraints,
relating the (space-like part of the) connection to the triad.
It should be investigated whether it would be possible to impose
them entirely, maybe using a triad representation
as done in the self-dual complex Ashtekar formulation to deal with
the reality conditions.

The covariant approach also opens the door to
study of Lorentz boosts and related covariance issues in Loop Quantum
Gravity, as already discussed in \cite{simone}.
Finally, the present analysis should be generalised to higher dimensions
and it should be possible to prove the same link at the kinematical level
between the canonical approach and the spin foam model as described in
\cite{coset}.

\section*{Acknowledgements}
The research of S.A. has been supported in part by European
network EUROGRID HPRN-CT-1999-00161.

\appendix

\section{Lorentz covariant canonical formulation}
\label{A}

In this appendix we
review the covariant formalism developed in \cite{sergei1}.
It is a canonical formulation of general relativity based on
the generalized Hilbert--Palatini action (\ref{palat}).
We use the following $3+1$ decomposition of the fields
\beq
&e^0=Ndt+\chi_a E_i^a dx^i ,\quad e^a=E^a_idx^i+E^a_iN^idt,& \nonumber \\
&{\tE}^i_a =h^{1/2}E^i_a,  \quad
\tN=h^{-1/2} N, \quad \sqrt{h}=\det E^a_i.& \label{tetrad}
\eeq
Here $E^i_a$ is the inverse of $E_i^a$.
The field $\chi_a$ describes deviation
of the normal to the hypersurface $\{ t=0\}$ from
the time direction.
Let us change the lapse and shift variables as follows
\be
N^i=\nd^i+\tE^i_a\chi^a\tNn, \quad
\tN=\tNn+\Et_i^a\chi_a\nd^i
\ee
and introduce
the multiplets which play the role of canonical variables
\beq
 & A_i^{ X}=(\omega_i^{0a},\frac12 {\eps^a}_{bc}\omega_i^{bc})
     &{\rm -\ connection\ multiplet}, \nonumber\\
&  \tP_X^{ i}=(\tE^i_a,{\eps_a}^{bc}\tE^i_b\chi_c)
     &{\rm -\ first\ triad\ multiplet}, \label{multHP}\\
&  \tQ_X^{ i}=(-{\eps_a}^{bc}\tE^i_b\chi_c,\tE^i_a)
     &{\rm -\ second\ triad\ multiplet}, \nonumber\\
&  \tPb_X^i=\tP_X^i-\frac{1}{\beta}\tQ_X^i
     &{\rm -\ canonical\ triad\ multiplet}. \nonumber
 \eeq
All triad multiplets are related by numerical matrices
\beq
&
\tP^i_X=\Pi_X^Y\tQ^i_Y, \qquad
\tP^i_X=\frac{\R_X^Y}{1+\frac{1}{\beta^2}}\tPb^i_Y,
& \\ &
g^{XY}=\left(
\begin{array}{cc}
1&0 \\ 0&-1
\end{array}
\right)\delta_a^b,  \qquad
\Pi^{XY} =\left(
\begin{array}{cc}
0&1 \\ 1&0
\end{array}
\right)\delta_a^b,  \qquad
\R^{XY} =g^{XY}-\frac{1}{\beta}\Pi^{XY}=\left(
\begin{array}{cc}
1& -\frac{1}{\beta} \\
 -\frac{1}{\beta} & -1
\end{array}
\right)\delta_a^b .  \label{mat}
&
\eeq
Also one can introduce the {\it inverse triad multiplets} $\Pt_i^X$ and
$\Qt_i^X$ and {\it projectors}
\beq
I_{(P)X}^Y &\equiv& \tP^{ i}_X\Pt_i^{Y}=\left(
\begin{array}{cc}
\frac{\delta_a^b-\chi_a\chi^b}{1-\chi^2} &
\frac{ {\eps_a}^{bc}\chi_c}{1-\chi^2} \\
\frac{ {\eps_a}^{bc}\chi_c}{1-\chi^2} &
-\frac{\delta_a^b\chi^2-\chi_a\chi^b}{1-\chi^2}
\end{array} \right),
\\
 I_{(Q)X}^Y &\equiv& \tQ^{ i}_X\Qt_i^{Y}=\left(
\begin{array}{cc}
-\frac{\delta_a^b\chi^2-\chi_a\chi^b}{1-\chi^2} &
-\frac{ {\eps_a}^{bc}\chi_c}{1-\chi^2} \\
-\frac{ {\eps_a}^{bc}\chi_c}{1-\chi^2} &
\frac{\delta_a^b-\chi_a\chi^b}{1-\chi^2}
\end{array} \right). \label{Qi-Qi}
\eeq
These projectors are functions of the field $\chi$ only and
can be considered respectively as projectors in the Lorentz algebra
on the $su(2)$ subalgebra defined by $\chi$
and its orthogonal boost part (see appendix B for more details).

The decomposed action reads
\beq
S_{(\beta)} &=&\int dt\, d^3 x (\tPb^i_X\partial_t A^X_i
+A_{0}^X \G_X+\nd^i H_i+\tNn H),  \label{dact} \\
\G_X&=&\partial_i \tPb^i_X +f_{XY}^Z A^Y_i \tPb^i_Z, \label{Gauss} \\
H_i&=&-\tPb^j_X F_{ij}^X,  \label{hdiff} \\
H&=&-\frac{1}{2\left(1+\frac{1}{\beta^2}\right)}
\tPb^i_X \tPb^j_Y f^{XY}_Z R^Z_W F_{ij}^W, \label{ham} \\
F^X_{ij}&=&\partial_i A_j^X-
\partial_j A_i^X+f_{YZ}^X A^Y_i A^Z_j. \label{FF}
\eeq

The action (\ref{dact}) gives rise to 10 first class constraints
$\G_X$, $H_i$, $H$ and also
two sets of second class constraints
\beq \phi^{ij}&=&\Pi^{XY}\tQ^i_X\tQ^j_Y=0, \label{phi} \\
\psi^{ij}&=&2f^{XYZ}\tQ_X^{l}\tQ_Y^{\{ j}\partial_l \tQ_Z^{i\} }
-2(\tQ\tQ)^{ ij }\tQ_Z^{l}A_l^Z+
2(\tQ\tQ)^{l\{i  }\tQ_Z^{j\}}A_l^Z=0, \label{psi} \\
{\rm where\ }(\tQ\tQ)^{ij}&=&g^{XY}\tQ_X^i\tQ_Y^j, \nonumber
\eeq
which lead to Dirac brackets.
They change the commutation relations of the canonical variables
so that the connection becomes non-commutative. Another
consequence is that the area operator is not diagonal
in the basis of Wilson lines defined with the connection $A_i^X$,
since the bracket $\{ A^X_i,\tP_Y^j\}_D$ is not diagonal in spatial indices.
The action of the area operator on such a Wilson line
depends on the details of embedding
of the surface and line into 3d space \cite{sergei2}.

However, one can redefine the connection in such a way that
the area operator would be diagonal on
Wilson lines defined with the new connection.
It has been shown  \cite{sergei3} that there is a two-parameter
family of such Lorentz connections, {\it i.e.}, transforming correctly
under the Gauss and diffeomorphism constraints.
They are written as
\beq
\SA_i^X(a,b) &=&
A_i^X+\frac{1}{2}\left( (1+\frac{a}{\beta})g^{XX'}
- \frac{1}{\beta}(1-b) \Pi^{XX'}\right)
I_{(Q)X'}^T\frac{R_T^Y}{1+\frac{1}{\beta^2}}f^W_{YZ}\Pt_i^Z \G_W
\nonumber \\
&+&
(a \delta^{X}_{X'} + b \Pi^{X}_{X'})
\left( I_{(Q)}^{X'W} \Pi_{WZ} A_i^Z + \ff_i^{X'}(\tQ)\right),
\label{consh}
\eeq
where we introduced
\be
\ff_i^{X}(\tQ)=g^{XX'} \Pi_{R}^Z\left(
I_{(Q)X'}^{R} f^{W}_{YZ} \Qt_i^Y \p_l \tQ^l_W
+ \frac12 f^{W}_{YZ} \tQ^k_{X'} \Qt_i^Y \tQ^l_W \p_k \Qt_l^R\right).
\label{fff}
\ee
Using the explicit expressions (\ref{multHP}), one can show that actually
$\ff^X_i(\tQ)$ depends only on the field $\chi$:
\be
\ff^X_i(\tQ)=-g^{XY}\frac{\tQ_X^j \Et_j^a \p_i\chi_a }{1-\chi^2}
=\left(- \frac{\eps^{abc}\chi_b\p_i\chi_c}{1-\chi^2},
\frac{\p_i\chi^a}{1-\chi^2} \right). \label{ff}
\ee
The connections (\ref{consh}) possess the following properties:
\beq
\{ \SA^X_i(a,b),\tP_Y^j\}_D&=&\delta_i^j \left(
(1-b)\delta^X_{X'} +a\Pi^X_{X'} \right) I_{(P)Y}^{X'},
\label{newAiP} \\
\{ \SA^X_i(a,b),I_{(P)}^{YZ}\}_D&=&0, \label{chicomm}\\
\{ \{ \SA_i^X(a,b), \SA_j^Y(a,b)\}_D , \tP^k_Z \}_D&=&0. \label{Jac}
\eeq
From (\ref{chicomm}) it follows that
\be
\{ \SA^X_i(a,b),\chi^a\}_D=0. \label{cAchi}
\ee
Note also that they can be represented as follows
\be
\SA_i^X(a,b) =
\SA_i^X+
(a \delta^{X}_{X'} + b \Pi^{X}_{X'})
\left( I_{(Q)}^{X'W} \Pi_{WZ} \SA_i^Z + \ff_i^{X'}(\tQ)\right),
\label{consh0}
\ee
where $\SA_i^X\equiv\SA_i^X(0,0)$ is the unique true
space-time connection diagonalizing the area operator \cite{sergei3}.
More precisely, it is the only connection from the family (\ref{consh}),
which transforms correctly under the time diffeomorphisms.

The resulting area spectrum for generic $(a,b)$ is
\be
{\cal S}\sim \hbar \sqrt{(a^2 + (1-b)^2) C(su(2)) -(1-b)^2 C_1(sl(2,C))
-a(1-b) C_2(sl(2,C))}
\label{spectab}
\ee
where $C_1=g^{XY}T_XT_Y$ and $C_2=\Pi^{XY}T_XT_Y$ are the Casimir operators
of $SL(2,C)$ and $C=I_{(Q)}^{XY}T_XT_Y$ is the Casimir of the $SU(2)$
subgroup obtained from the canonically embedded one by a boost
transformation with the parameter $\chi^a$.

\section{Reducing \SL to the subgroup \sgchi}

\label{C}

Let us introduce mixed tensors:
\beq
 & q_X^a=\left({\eps^{a}}_{bc}\chi^c,\delta^a_b\right), \qquad
p_X^a=\left(\delta_b^a,-{\eps^{a}}_{bc}\chi^c\right), & \\
 & q^X_a=\left(-\frac{{\eps_{a}}^{bc}\chi_c}{1-\chi^2},
\frac{\delta_a^b-\chi_a\chi^b}{1-\chi^2}\right), \qquad
  p^X_a=\left(\frac{\delta_a^b-\chi_a\chi^b}{1-\chi^2},
\frac{{\eps_{a}}^{bc}\chi_c}{1-\chi^2}\right). &
\eeq
They relate the triad and the triad multiplets
\beq
& \tQ^i_X=q^a_X\tE^i_a, \qquad \tP^i_X=p^a_X\tE^i_a, & \\
& \Qt_i^X=q_a^X\Et_i^a, \qquad \Pt_i^X=p_a^X\Et_i^a &
\eeq
and possess the following properties:
\beq
& q_X^a p^X_b=p_X^a q^X_b=0, & \\
& q_X^a q^X_b=p_X^a p^X_b=\delta^a_b, & \\
& q_X^a q^Y_a=I_{(Q)X}^Y, \quad p_X^a p^Y_a=I_{(P)X}^Y. &
\eeq
One can say that the set of 6-dimensional vectors $(p^a,q^a)$ describes
a basis in the Lorentz algebra obtained from the standard one by a
Lorentz boost with parameter $\chi$.

It is trivial to show that the quantity (\ref{AshBar})
satisfy $p_X^a \iAash{i}{X}=0$. Then one can show that the
remaining components of the connection coincide with
the quantity \Ref{connecbarros}:
\be
\iAas{i}{a}\equiv q_X^a \iAash{i}{X}=\beta \iAbar{i}{a}.
\label{nonash}
\ee
Thus, $\iAash{i}{X}$ gives an embedding of $\iAbar{i}{a}$ into the
Lorentz algebra.

To proceed further, we introduce the "boosted" \sg subgroup
(let us call it \sgchi) with generators
\be
\Gc_a=q_a^X\G_X.
\ee
They form $su(2)$ algebra with the structure constants
$f_{ab}^c=-{\eps_{abd}}
\left(\delta^{dc}+\frac{\chi^d\chi^c}{1-\chi^2}\right)$:
$\{\Gc_a,\Gc_b\}_D=f_{ab}^c\Gc_c$.\footnote{In the case of constant
$\chi$, one can redefine
the generators by a matrix constructed from $\chi$ to get the
algebra with the usual structure constants ${\eps_{ab}}^c$.}
It turns out that if $\chi=const$, $\iAas{i}{a}$
transforms as a connection with respect to
the transformations generated by $\Gc_a$:
\be
\{ \Gc(n), \iAas{i}{a}\}_D=\partial_i n^a +f^a_{bc}n^b \iAas{i}{c}.
\label{trash}
\ee
As a result, in this restricted situation, $\iAas{i}{a}$ is a
connection of the gauge group \sgchi and generalizes
the Ashtekar--Barbero connection.
Similarly, the restriction of the Lorentz covariant connection
$\SSA_i^X$ (\ref{conSU2})
to \sgchi given by
\be
\iSSAr{i}{a}=q_X^a\SSA_i^X  \label{iSSA}
\ee
transforms according to (\ref{trash}) since $\Gc_a$ commute with $\chi$.

\section{Commutator of two connections}
\label{B}

The commutator of the space-time connection $\SA_i^a$ with itself
can be calculated and is given by a horrible expression \cite{sergei4}:
\beq
&\{ \int d^3x\, f(x)\SA^X_i(x), \int d^3y\, g(y)\SA^Y_j(y)\}_D=&
\nonumber \\
& \frac{1}{2\left(1+\frac{1}{\beta^2}\right)} R^X_S R^Y_T \int d^3z\, \left[
\left(  K^{ST,l}_{ij} g\p_l f- K^{TS,l}_{ji} f\p_l g \right) +
fg \left( L^{ST}_{ij}-L^{TS}_{ji} \right)
\right], &
\label{AAcom}
\eeq
where
\beq
K^{ST,l}_{ij}&=& \Pi^{SS'}f_{S'}^{PQ}\left[\tQ^l_P\left(
(\Qt\Qt)_{ij}\IQ^T_Q+\Qt_i^T\Qt_j^Q-\Qt_j^T\Qt_i^Q\right)
+\delta_i^l\IQ^T_Q\Qt_j^P\right],
\nonumber \\
L^{ST}_{ij}&=&
\Pi^{S}_{S'}f^{PQ}_Z \left[ \Qt_j^{S'}\Qt_n^T\Qt_i^Z+
(\Qt\Qt)_{in}\Qt_j^{S'}\IQ^{TZ}+\Qt_i^T\Qt_n^{S'}\Qt_j^Z
\right. \nonumber \\
&& \quad \qquad  - \left.
\Qt_i^T\Qt_j^{S'}\Qt_n^Z+(\Qt\Qt)_{ij}\Qt_n^{S'}\IQ^{TZ}-
\Qt_j^T\Qt_n^{S'}\Qt_i^Z\right] \tQ^l_P\p_l\tQ^n_Q
\nonumber \\
&+& \Pi^{S}_{S'}f^{Q}_{ZP} \left[ \Qt_n^T\Qt_j^P+
(\Qt\Qt)_{jn}\IQ^{TP}-\Qt_j^T\Qt_n^{P}\right]
\IQ^{ZS'}\p_i\tQ^n_Q
\nonumber \\
&+& \Pi_{Z}^{Z'}f^{PQ}_{Z'}\left[
(\Qt\Qt)_{in}\Qt_j^{Z}\IQ^{ST}-(\Qt\Qt)_{in}\Qt_j^T\IQ^{SZ}-
(\Qt\Qt)_{ij}\Qt_n^T\IQ^{SZ}  \right]  \tQ^l_P\p_l\tQ^n_Q
\nonumber \\
&+& \Pi^S_{S'}f^Z_{PQ}\Qt_j^{S'}\Qt_i^Q\IQ^{TP}\p_l\tQ_{Z}^l+
f^Z_{PQ}\Qt_i^P\Qt_j^Q\IQ^T_Z\IQ^{SW}\Pi_W^{W'}\p_l\tQ_{W'}^l.
\eeq

Nevertheless, using this expression one can obtain several important
properties. First of all, note that
from (\ref{AAcom}) it is straightforward to check the following fact:
\be
I_{(P)S}^X (R^{-1})^S_Z\{ \SA_i^Z,\SA_j^W\}_D (R^{-1})^T_W I_{(P)T}^Y=0.
\label{IPAIPA}
\ee
Then, since the quantity (\ref{AshBar}) can be written as a similar
projection of the space-time connection $\SA_i^a$
\be
\iAash{i}{X}
=-\beta\left(1+\f{1}{\beta^2}\right)\Pi^X_Y I_{(P)T}^Y (R^{-1})^T_Z\SA_i^Z,
\ee
it commutes with itself. Moreover, from equation
\Ref{cAchi} it follows that any $\SA(a,b)$  commutes with
$\chi$ and, therefore, with the function $\ff(\tl{Q})$ (\ref{ff}).
This allows to conclude that the covariant generalization of
the Ashtekar--Barbero connection $\SSA$ given by (\ref{conSU2})
is also commuting.

Another property, which follows immediately from the relation
\Ref{contA}, is
\be
I_{(Q)Z}^X \{ \SA_i^Z,\SA_j^W\}_D I_{(Q)W}^Y=0.
\label{IAIA}
\ee

\section{Representations of $SL(2,C)$}

\label{D}

The generators $T_X$ form the $sl(2,C)$ algebra with the structure constants
$f_{XY}^Z$:
\be
[T_X, T_Y]=f_{XY}^Z T_Z.
\ee
Let us introduce the notations $T_X=(A_a,-B_a)$ and
\beq
&H_+=iB_1-B_2, \qquad H_-=iB_1+B_2, \qquad H_3=iB_3,& \\
&F_+=iA_1-A_2, \qquad F_-=iA_1+A_2, \qquad F_3=iA_3.&
\eeq
These generators commute in the following way:
\beq
& [H_+,H_3]=-H_+, \qquad [H_-,H_3]=H_-, \qquad [H_+,H_-]=2H_3, &
\nonumber \\
& [H_+,F_+]=[H_-,F_-]=[H_3,F_3]=0, & \nonumber \\
& [H_+,F_3]=-F_+, \qquad  [H_-,F_3]=F_-, & \\
& [H_+,F_-]=-[H_-,F_+]=2F_3, & \nonumber \\
& [F_+,H_3]=-F_+, \qquad [F_-,H_3]=F_-, & \nonumber \\
& [F_+,F_3]=H_+, \qquad [F_-,F_3]=-H_-, \qquad [F_+,F_-]=-2H_3. &
\nonumber
\eeq

An irreducible representation of the Lorentz group is characterized
by two numbers $(n,\mu)$, where $n\in \Nat/2$ and $\mu\in \Cb$.
In the space $\CH_{n,\mu}$ of this representation one can introduce
an orthonormal basis
\be
\{ \xi_{j,m}\},\qquad m=-j,-j+1,\dots,j-1,j, \quad
l=n,n+1,\dots
\ee
such that the generators introduced above act in the
following way \cite{GMS}:
\beq
H_3\xi_{j,m}&=& m\xi_{j,m}, \nonumber \\
H_+\xi_{j,m}&=& \sqrt{(j+m+1)(j-m)}\xi_{j,m+1}, \label{gauss-rep} \\
H_-\xi_{j,m}&=& \sqrt{(j+m)(j-m+1)}\xi_{j,m-1}, \nonumber \\
F_3\xi_{j,m}&=& \gamma_{(j)}\sqrt{j^2-m^2}\xi_{j-1,m}+\beta_{(j)}m\xi_{j,m}
-\gamma_{(j+1)}\sqrt{(j+1)^2-m^2}\xi_{j+1,m}, \nonumber \\
F_+\xi_{j,m}&=&
\gamma_{(j)}\sqrt{(j-m)(j-m-1)}\xi_{j-1,m+1}+\beta_{(j)}\sqrt{(j-m)(j+m+1)}
\xi_{j,m+1} \nonumber \\
&+& \gamma_{(j+1)}\sqrt{(j+m+1)(j+m+2)}\xi_{j+1,m+1}, \label{boosts-rep} \\
F_-\xi_{j,m}&=&
-\gamma_{(j)}\sqrt{(j+m)(j+m-1)}\xi_{j-1,m-1}+\beta_{(j)}\sqrt{(j+m)(j-m+1)}
\xi_{j,m-1} \nonumber \\
&-& \gamma_{(j+1)}\sqrt{(j-m+1)(j-m+2)}\xi_{j+1,m-1}, \nonumber
\eeq
where
\be
\beta_{(j)}=-\frac{in\mu}{j(j+1)}, \qquad
\gamma_{(j)}=\frac{i}{2j}
\sqrt{\frac{(j^2-n^2)(j^2-\mu^2)}{
\left(j-\f{1}{2}\right)
\left(j+\f{1}{2}\right)
}}.
\ee
The unitary representations correspond to two cases:
\beq
1)& (n,\mu)=(n,i\rho),  \quad n\in \Nat/2, \ \rho\in \Rb & \quad - \
principal\ series, \\
2)& (n,\mu)=(0,\rho), \quad |\rho|<1, \ \rho\in \Rb & \quad - \
supplementary\ series.
\eeq
The principal series representations are the ones intervening
in the Plancherel decomposition formula for $L^2$ functions over $SL(2,C)$.
Simple representations are the representations of the principal series
with the vanishing Casimir  $C_2(sl(2,C))$ \cite{bc1,bc2}.
There are two types of such representations: $(n,0)$ and $(0,i\rho)$.
In both cases we have $\beta_{(j)}=0$ for all $j$.
However, the representations $(0,i\rho)$ have the particularity
that they possess an $SU(2)$ invariant vector $\xi_{0,0}$.

\section{Creating a Lorentz connection from an \SU connection}
\label{so3toso31}

Let us suppose that we have a hypersurface with an \SU connection $a$.
Now, we would like to extend the \SU structure to a \SL one and
the connection $a$ to a \SL connection $A$.
For this purpose, we introduce on the hypersurface a time normal
field $\chi\in{\cal H}_+=SL(2,C)/SU(2)$
valued on the (upper part of the two-sheet) hyperboloid.
We choose a reference vector $\chi_0$ on the hyperboloid and
define $x\in SL(2,C)$ rotating from $\chi_0$ to $\chi$.
Let us define the isomorphism
\be
i_{\chi}:su(2)\rightarrow su_{\chi}(2)
\ee
and the projection
\be
I_{\chi}:sl(2,C)\rightarrow su_{\chi}(2).
\ee
Then, we would like to
define the Lorentz connection so that $I_{\chi}A=i_{\chi}a$.
Let  $A=i_{\chi}a+b$.
Action of a gauge transformation $g\in SL(2,C)$
is defined by splitting $g$ into its $SU_\chi(2)$ part acting with the
initial \SU action on $a$ and its pure boost part
acting on the field $\chi$.
Good behaviour of $A$ under gauge transformation gives
the following conditions on $b$:
\be
\forall g\in SU_{\chi}(2),\,
{}^gb=gbg^{-1}+x(\p x^{-1}) +g(\p x)x^{-1}g^{-1}
\ee
and
\be
\forall g \textrm{ pure boost},\,
{}^gb=gbg^{-1}+g(\p g^{-1}).
\ee
A priori, there are many solutions to these equations.
The simplest solution is $b=x(\p x^{-1})$.
Thus, this procedure allows to create an \SL
connection from an \SU one by introducing the vector field $\chi$.
It can be used to construct the covariant generalization of the
Ashtekar-Barbero connection \Ref{conSU2}.
Indeed, one can start from the $SU(2)$ Loop Gravity, with a hypersurface
provided with an \SU connection $a$ (the Ashtekar-Barbero connection),
and, following this procedure, one can try to make the formalism
explicitly covariant. First, we introduce a time normal field $\chi$
(which allows to go out of the time gauge) and then we construct
$A=i_{\chi}a+x(\p x^{-1})$, which gives rise to an $SL(2,C)$
structure with a covariant
connection on the canonical hypersurface.
It turns out that this result matches exactly what comes
out of the direct covariant canonical analysis as described in section 2:
$b=x(\p x^{-1})$ coincides with the function $\ff(\tQ)$
and $i_{\chi}a$ should be identified with $\SSAr$ \Ref{iSSA}.

\end{document}